\documentclass[pra,10pt,notitlepage,twocolumn]{revtex4-1}

\usepackage{graphicx}

\usepackage{dcolumn}

\usepackage{bm}

\usepackage{longtable}

\usepackage{graphics}

\usepackage{amssymb}

\usepackage{amsmath}

\usepackage{xspace}

\usepackage{epsfig}

\usepackage{rotating}

\usepackage{epstopdf}
\usepackage{hyperref}

\newcommand{\etal}{\emph{et al}.\xspace}

\DeclareFontFamily{U}{rcjhbltx}{}
\DeclareFontShape{U}{rcjhbltx}{m}{n}{<->rcjhbltx}{}
\DeclareSymbolFont{hebrewletters}{U}{rcjhbltx}{m}{n}

\let\aleph\relax\let\beth\relax
\let\gimel\relax\let\daleth\relax

\DeclareMathSymbol{\aleph}{\mathord}{hebrewletters}{39}
\DeclareMathSymbol{\beth}{\mathord}{hebrewletters}{98}\let\bet\beth
\DeclareMathSymbol{\gimel}{\mathord}{hebrewletters}{103}
\DeclareMathSymbol{\daleth}{\mathord}{hebrewletters}{100}

\DeclareMathSymbol{\lamed}{\mathord}{hebrewletters}{108}
\DeclareMathSymbol{\mem}{\mathord}{hebrewletters}{109}
\DeclareMathSymbol{\ayin}{\mathord}{hebrewletters}{96}
\DeclareMathSymbol{\tsadi}{\mathord}{hebrewletters}{118}
\DeclareMathSymbol{\qof}{\mathord}{hebrewletters}{114}
\DeclareMathSymbol{\shin}{\mathord}{hebrewletters}{152}
\DeclareMathSymbol{\resh}{\mathord}{hebrewletters}{114}
\DeclareMathSymbol{\tav}{\mathord}{hebrewletters}{116}

\begin{document}
	
	\title{Topological frustration and structural balance in strongly correlated itinerant electron systems: an extension of Nagaoka's theorem}

	\author{B. J. Powell}\affiliation{School of Mathematics and Physics, The University of Queensland, Brisbane, Queensland, 4072, Australia}
	
	\begin{abstract}
	We prove that Nagaoka's theorem, that the large-$U$ Hubbard model with exactly one hole is ferromagnetic, holds for any balanced Hamiltonian. Simply put, if a positive bond encodes friendship and a negative bond encodes enmity, then balance implies when that enemy of one's enemy is one's friend. We argue that, in itinerant electron systems, a balanced Hamiltonian, rather than bipartite lattice, defines an unfrustrated system. The proof is valid for multi-orbital models with arbitrary two-orbital interactions provided that no exchange interactions are antiferromagnetic: a class of models  including the Kanomori Hamiltonian.  
	\end{abstract}
	
	\maketitle
%
%
%
%
%

%

\section{Introduction}

Geometrical frustration plays a central role in the understanding of quantum magnetism \cite{frust}. The distinction between bipartite and geometrically frustrated lattices is fundamental for spin-models of localized electrons. But, we will argue below, in itinerant electron systems the bipartite/geometrically frustrated distinction is not relevant. Rather a topological definition of frustration is required. We will argue below that structural balance, introduced  in an anthropological context, where one is often concerned with networks containing both beneficial and antagonistic relationships, \cite{Hage}, provides an appropriate measure of topological frustration. 

Previous attempts to classify the frustration of itinerant electrons have focused on the reduced bandwidth in frustrated systems \cite{Barford,Merino}. Therefore, these measures miss the fundamental role that the sign of the hopping integral plays in frustrated itinerant electron systems. Balance considers this.

To ground the above claims we study one of the few exact results of strongly correlated itinerant electrons: Nagaoka's theorem. This theorem concerns the properties of the infinite-$U$ Hubbard model -- however, here we consider a larger class of Hamiltonians, which allows for arbitrary two-orbital interactions:
\begin{eqnarray}
\hat{\cal H}&=&\hat{\cal H}_t+\hat{\cal H}_U+\hat{\cal H}_V+\hat{\cal H}_J+\hat{\cal H}_{P}+\hat{\cal H}_X \label{Eq:full}
\end{eqnarray}
where 
\begin{subequations}
	\begin{eqnarray}
	\hat{\cal H}_t&=&-\sum_{ij\mu\nu\sigma}t_{ij\mu\nu}\hat c^\dagger_{i\mu\sigma}\hat c_{j\nu\sigma}, \label{Eq:tb}\\
	\hat{\cal H}_U&=&U\sum_{i\mu} \hat n_{i\mu\uparrow} \hat n_{i\mu\downarrow},\\
	\hat{\cal H}_V&=&\hat V(\{\hat n_{i\mu}\}),\\
	\hat{\cal H}_J&=&\sum_{(i,\mu)\ne(j,\nu)} \sum_{\sigma\rho} J_{ij\mu\nu}\hat c^\dagger_{i\mu\sigma}\hat c^\dagger_{j\nu\rho}\hat c_{i\mu\rho}\hat c_{j\nu\sigma}
	\notag\\	&=&
	\sum_{(i,\mu)\ne(j,\nu)} J_{ij\mu\nu}\left(\hat{\bm{S}}_{i\mu}\cdot\hat{\bm{S}}_{j\nu}+\frac14 \hat{n}_{i\mu}\hat{n}_{j\nu}\right),\\
	\hat{\cal H}_{P}&=&\sum_{(i,\mu)\ne(j,\nu)}P_{ij\mu\nu}\hat c^\dagger_{i\mu\uparrow}\hat c^\dagger_{i\mu\downarrow}\hat c_{j\nu\uparrow}\hat c_{j\nu\downarrow},
	\end{eqnarray}
	\begin{eqnarray}
	\hat{\cal H}_X&=&\sum_{(i,\mu)\ne(j,\nu)}\sum_{\sigma}X_{ij\mu\nu}\hat c^\dagger_{i\mu\sigma}\hat c_{j\nu\sigma}(\hat n_{i\mu\overline\sigma}+\hat n_{j\nu\overline\sigma}-1),\hspace*{15pt}
	\end{eqnarray}
\end{subequations}
$\hat c_{i\mu\sigma}^{(\dag)}$ creates (annihilates) an electron with spin $\sigma$ on the $\mu$th Wannier orbital centered on site $i$, $\overline{\sigma}\ne\sigma$, $\hat n_{i\mu\sigma}=\hat c^\dagger_{i\mu\sigma}\hat c_{i\mu\sigma}$, $\hat n_{i\mu}=\hat n_{i\mu\uparrow}+\hat n_{i\mu\downarrow}$,  $\hat{\bm{S}}_{i\mu}=\frac12\sum_{\alpha\beta}\hat c^\dagger_{i\mu\alpha}{\bm\sigma}_{\alpha\beta}\hat c_{i\mu\beta}$, and ${\bm\sigma}$ is the vector of Pauli matrices.
Here $t_{ij\mu\nu}$ is the amplitude for hopping between the $\mu$th orbital on site $i$ and the $\nu$th orbital on site $j$ \cite{foot-signs}, $U$ is the effective Coulomb interaction between two electrons on the same orbital, $\hat V(\{\hat n_{i\mu}\})$ is any potential that depends only on the orbital occupation numbers -- obviously this includes arbitrary pairwise direct Coulomb interaction: $\sum_{ij\mu\nu}V_{ij\mu\nu}\hat n_{i\mu}\hat n_{j\nu}$; but in fact the proofs below hold for arbitrary terms of this form, including multi-site interactions, the $J_{ij\mu\nu}$ are the exchange couplings, the $P_{ij\mu\nu}$ are pair hopping amplitudes, and the $X_{ij\mu\nu}$ are correlated hopping amplitudes.
Note that no assumption about the intra-site hopping has been made, in particular $t_{ii\mu\nu}$ may be non-zero.

It will be convenient below to differentiate between four versions of this model: (a) the multiorbital model -- Eq. (\ref{Eq:full}); (b) the single orbital model -- one orbital per site (henceforth we drop the orbital subscripts when discussing single orbital models) (c) the extended Hubbard model -- the single orbital model with $J_{ij}=P_{ij}=X_{ij}=0$ for all $i,j,\mu,\nu$; and (d) the Hubbard model -- the extended Hubbard model with $\hat V(\{\hat n_{i\mu}\})=0$.

Note that the hole doped sector of each of these models ($N<L$, where $N$ is the number of electrons and $L$ is the number of orbitals on the entire lattice) is equivalent to  the electron sector of that model ($N>L$) with the signs of all $t_{ij\mu\nu}$ reversed. A particle-hole transformation maps between the Hamiltonians, up to  constants, even in the absence of particle-hole symmetry. Henceforth we will only discuss the hole doped problem; however it is implicit throughout that all results  hold for the electron doped problem if the signs of all $t_{ij\mu\nu}$  reversed.

Nagaoka showed \cite{Nagaoka} that in the Nagaoka limit ($U\rightarrow\infty$ and $N=L-1$) the ground state of the Hubbard model on certain lattices is a fully polarized ferromagnet -- i.e., the magnetization, $M=N/2$.  Nagaoka showed that for nearest neighbor hopping only ($t_{ij}=t$ for nearest neighbors, $t_{ij}=0$ otherwise) and $t<0$ \cite{foot-signs} this result holds for simple cubic, body centered cubic, face centered cubic and hexagonal close packed lattices.

However, for $t>0$ Nagaoka found that the theorem holds on the simple cubic and body centered cubic lattices, but not for the face centered cubic or hexagonal close packed lattices. The former lattices are bipartite, while the latter are not. On a bipartite lattice the gauge transformation $\hat c_{i\sigma}\rightarrow-\hat c_{i\sigma}$ on one sublattice only changes the sign of all hopping integrals.

In 1989 Tasaki gave a more general proof of Nagaoka's theorem  \cite{Tasaki}. Specifically, Tasaki proved that Nagaoka's theorem holds for the extended Hubbard model on all lattices where  $t_{ij}\leq0$ for all $i,j$.
Thus, one is moved to ask which other Hamiltonians with some or all  $t_{ij}>0$ have ferromagnetic ground states? In particular, for which geometrically frustrated (non-bipartite) Hamiltonians can one prove Nagoaka's theorem? This is particularly important as for a simple covalent bonds one  expects that $t_{ij}>0$ \cite{Powell}.

In 1996 Kollar, Strack and Vollhardt \cite{Vollhardt} extend Nagaoka's theorem in a different direction --  discussing the role of other two-body interactions. Among other things they showed that Nagaoka's theorem holds for the infinite-$U$ single orbital model on periodic lattices if the hopping and all interactions are constrained to be between nearest neighbors only, exchange is ferromagnetic (or zero), and either $t<0$ or the lattice is bipartite. It is therefore natural to ask what other (e.g., longer range) two-orbital interactions admit that a proof of Nagaoka's theorem?

Furthermore, given that ferromagnetism is observed in many materials where multiple orbitals are relevant to the low-energy physics it is natural to ask whether multiple orbital models exhibit Nagaoka-like ferromagnetism.

Below, we give partial answers to the  above questions by proving the following:

%
%

\textit{Theorem 1}: Consider the multiorbital model (\ref{Eq:full}) with $U$ infinite, $V$ and $\{P_{ij\mu\nu}\}$  arbitrary,  $J_{ij\mu\nu}\leq0$ for all $i,j,\mu,\nu$ and $N=L-1$. If the signed graph $\mathcal{S}$ defined by 
the set of renormalized hopping integrals $\{t^*_{ij\mu\nu}\}$, where $t^*_{ij\mu\nu}\equiv t_{ij\mu\nu}+X_{ij\mu\nu}$, is balanced (defined below) then among the ground states there exist at least $L$ states with $S=S_\text{max}\equiv N/2$.

\textit{Theorem 2}: Consider the multiorbital model (\ref{Eq:full}) with $U$ infinite, $V$ and $\{P_{ij\mu\nu}\}$  arbitrary,  $J_{ij\mu\nu}\leq0$ for all $i,j,\mu,\nu$ and $N=L-1$. If the signed graph $\mathcal{S}$ defined by 
the set of renormalized hopping integrals $\{t^*_{ij\mu\nu}\}$ is balanced and $\hat{\cal H}_t+\hat{\cal H}_X$ satisfies the connectivity condition (defined below) then the ground state has  $S=S_\text{max}\equiv N/2$ and is unique up to the trivial $N$-fold  degeneracy.

The remainder of the paper is laid out as follows. Having introduced balanced and unbalanced lattices in section \ref{sect:balance}, we prove that balance is a sufficient condition for the proof of Nagaoka's theorem in the Nagaoka limit (section \ref{sect:sufficient}). All of the lattices for which Nagaoka, Tasaki or Kollar \etal proved Nagaoka's theorem previously are balanced. 
Finally, in section \ref{sec:orbital} discuss the implications of balance for the orbital part of the ground state wavefunction, clarifying why balance favours ferromagnetism.

\section{Balance}\label{sect:balance}

The sign of the hopping between a  pair of orbitals, $t^*_{ij\mu\nu}$, is not gauge invariant: the  transformation $\hat c_{j\nu\sigma}\rightarrow e^{-i\theta_{j\nu}}\hat c_{j\nu\sigma}$ takes $t^*_{ij\mu\nu}\rightarrow t^*_{ij\mu\nu}e^{i\theta_{j\nu}}$. 
Nevertheless, gauge invariant  information is contained in the signs of the set $\{t^*_{ij\mu\nu}\}$ associated with a particular Hamiltonian. 

This is a topological problem -- the magnitudes of the $t^*_{ij\mu\nu}$ are unimportant; only their signs matter. 
Thus, rather than considering every  $\{t^*_{ij\mu\nu}\}$ separately, it is sufficient to instead study a related   `signed graph'. 
We define this signed graph as follows: We introduce a vertex of the graph  corresponding to each orbital in the physical Hamiltonian, $\{\aleph\}=\{(i,\mu)\}$ (throughout this paper we use Latin characters for sites in the Hamiltonian, Greek for orbitals, and Hebrew for vertices in the signed graph). Furthermore, we introduce a set of edges defined by $\{\tau_{(i,\mu),(j,\nu)}\}$, where $\tau_{(i,\mu),(j,\nu)}\equiv -\text{sgn}(t^*_{ij\mu\nu})$ if and only if $t^*_{ij\mu\nu}\ne0$.
We now ask whether there exists a series of gauge transformations that make all  $\tau_{\aleph\bet}=1$? If so, the gauge transformation  makes all  $t^*_{ij\mu\nu}\leq0$

\begin{figure}
	\begin{center}
		\includegraphics[width=0.32\columnwidth]{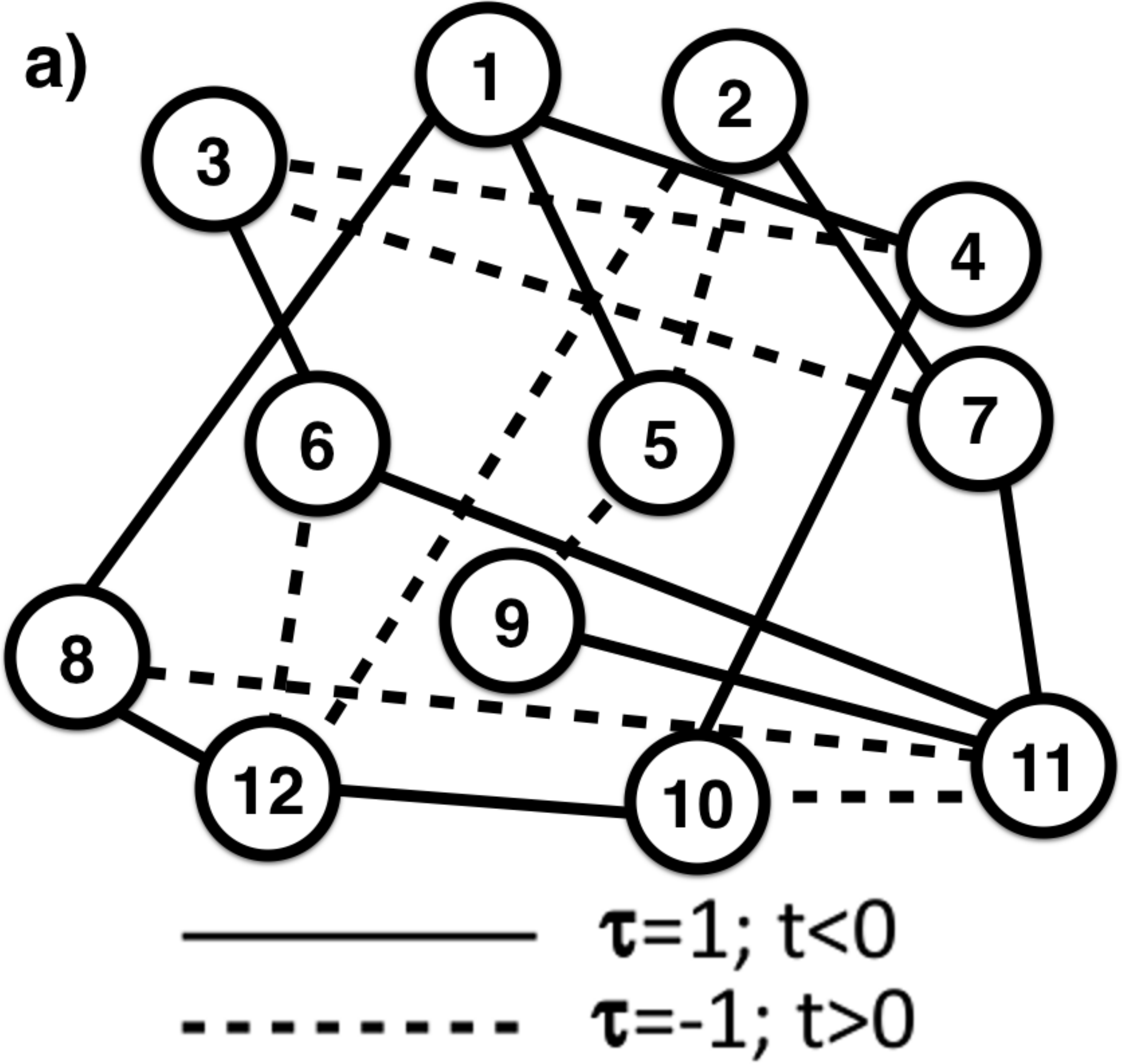}
		\includegraphics[width=0.32\columnwidth]{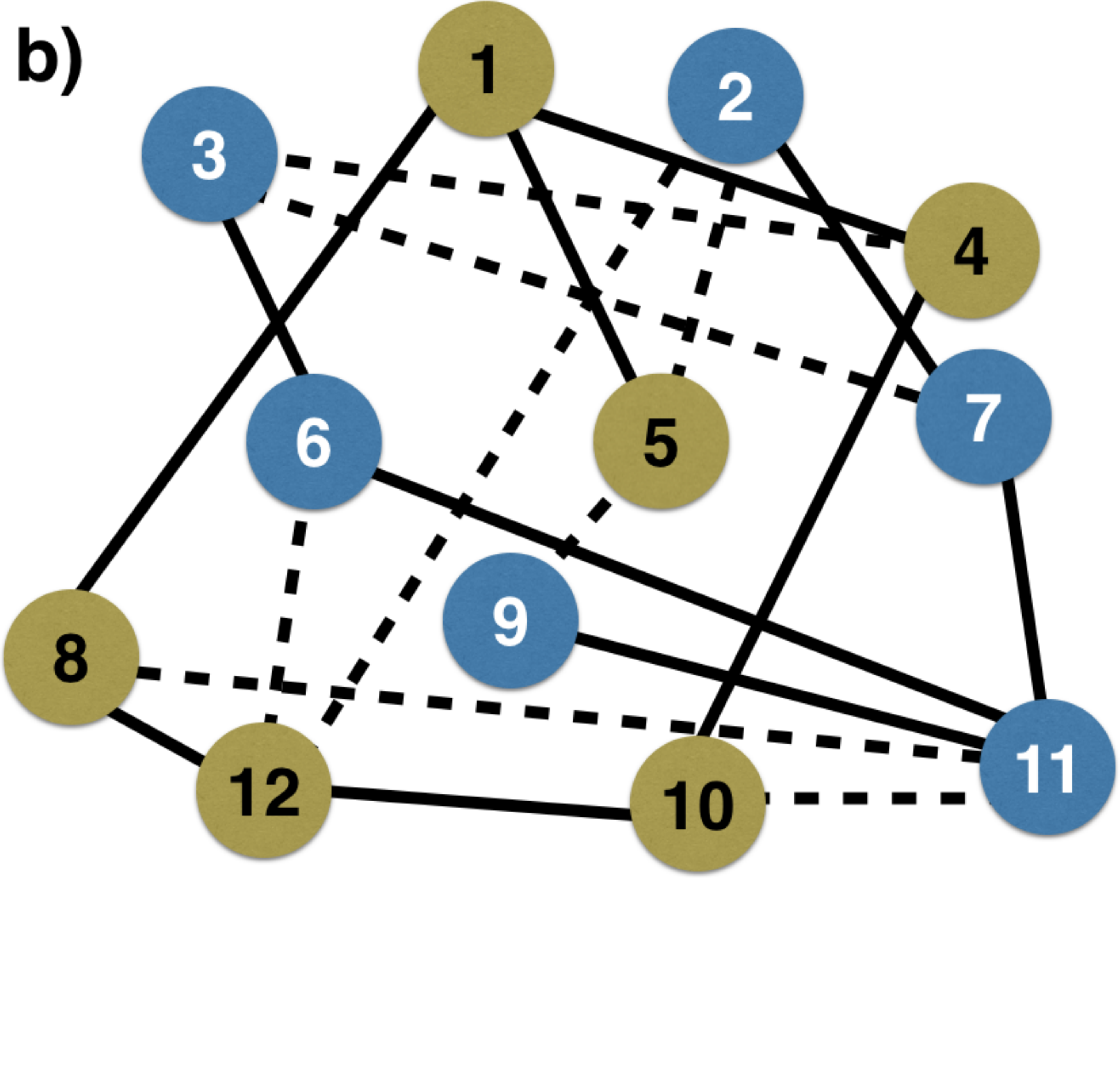}
		\includegraphics[width=0.32\columnwidth]{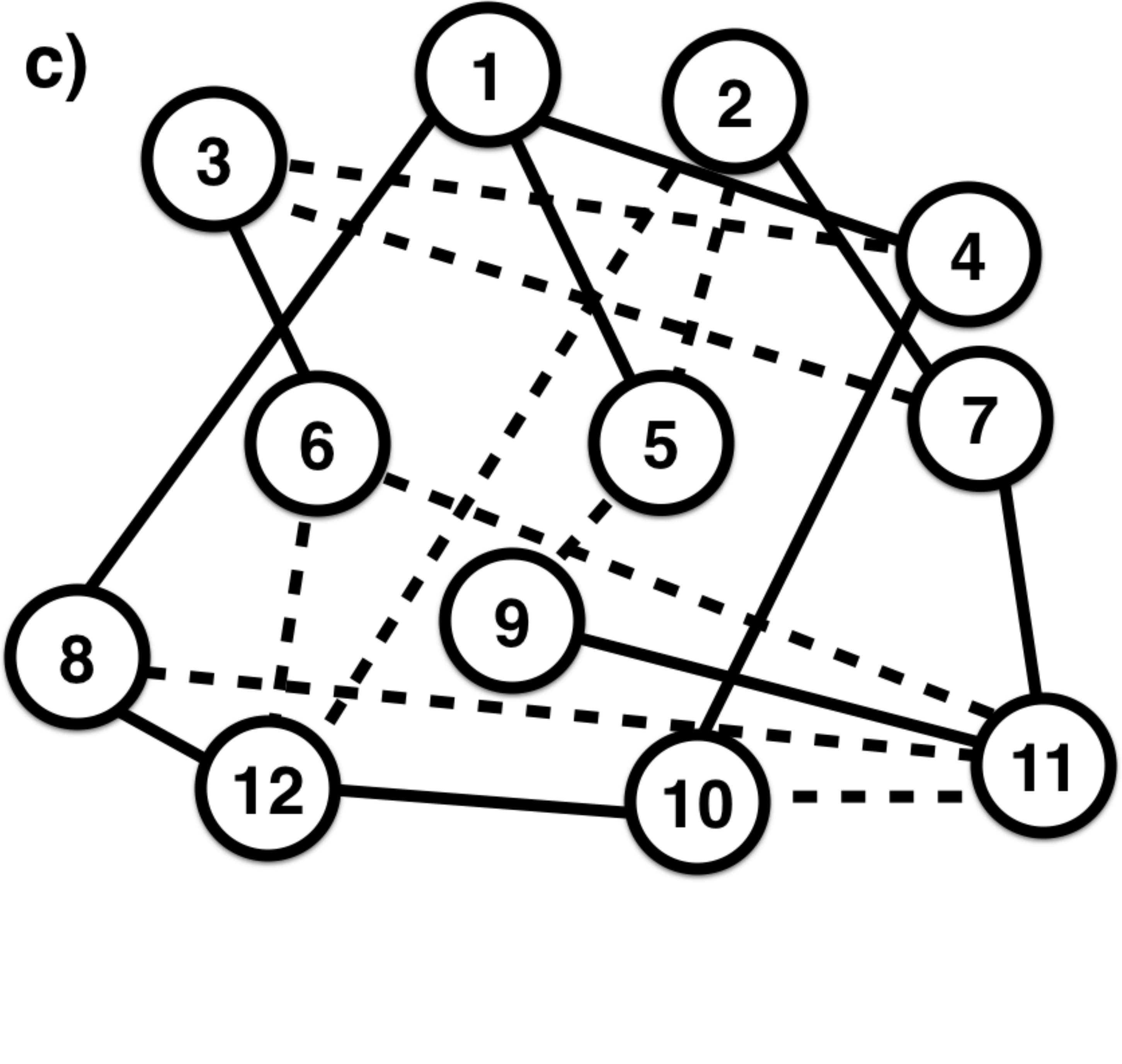}
	\end{center}
	\caption{Balanced and unbalanced signed graphs. In all panels solid lines indicate $\tau_{\aleph\bet}=1$ [$t_{ij\mu\nu}<0$, where $\aleph=(i,\mu)$ and $\bet=(j,\nu)$] and dashed lines indicate $\tau_{\aleph\bet}=-1$ ($t_{ij\mu\nu}>0$). (a) A balanced lattice. (b) As (a) with a choice of the subsets ${\mathcal S}_a$ and  ${\mathcal S}_b$ indicated by the shading of the vertices. (c) An unbalanced lattice: the path $6\rightarrow 11\rightarrow 10\rightarrow 12\rightarrow 6$ is negative ($\tau_{6,11}\tau_{11,10}\tau_{10,12}\tau_{12,6}=-1$).} 
	\label{fig}
\end{figure}

A \textit{walk} on a signed graph is defined as a sequence of vertices such that consecutive vertices in the sequence are connected by an edge, e.g., $\aleph\rightarrow\bet\rightarrow\gimel\rightarrow\dots\rightarrow\resh\rightarrow\shin\rightarrow\tav$.
A walk in which all the sites that are visited are distinct (i.e., a self avoiding walk) is called a \textit{path}.
A path that visits at least three sites and is closed (e.g., $\aleph=\tav$, in the example above) is called a \textit{cycle}.
The sign of a path or cycle on a signed graph is defined as the product of signs $\tau_{\aleph\bet} \tau_{\bet\gimel}\dots \tau_{\resh\shin}\tau_{\shin\tav}$) of the edges forming the path/cycle.
Thus every positive cycle has an even number (including zero) of negative edges.
A signed graph is  \textit{balanced} if all cycles in the corresponding signed graph are positive, cf. Fig. \ref{fig}.
We will call a Hamiltonian balanced if it maps onto a balanced signed graph.

The fundamental theorem of signed graphs \cite{Harary53} states that for a signed graph, $\cal S$, the following three conditions are equivalent:

\begin{enumerate}
	\item $\cal S$ is balanced, i.e., all cycles within $\cal S$ are positive.
	\item For any pair of vertices  $\aleph$ and $\bet$ in $\cal S$ all paths joining $\aleph$ and $\bet$ have the same sign.
	\item There exists a partition of $\mathcal S$ into two subsets, ${\mathcal S}_a$ and  ${\mathcal S}_b$, (one of which may be empty) such that  $\tau_{\aleph\bet}\geq0$ for all $\aleph$ and $\bet$ within the same subset, but   $\tau_{\aleph \bet}=\tau_{\bet\aleph}\leq0$ for $\aleph\in{\mathcal S}_a$ and $\bet\in{\mathcal S}_b$. \label{2sets}
\end{enumerate}

\begin{figure}
	\begin{center}
		\includegraphics[width=\columnwidth]{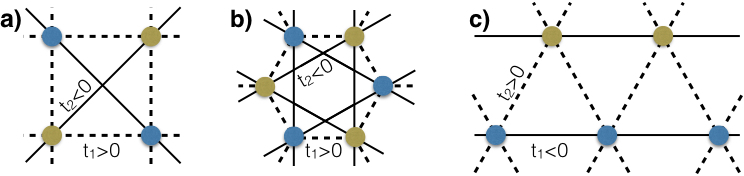}
	\end{center}
	\caption{Three geometrically frustrated but balanced periodic lattices. (a) The square lattice with nearest neighbour hopping $t_1>0$ and next nearest neighbour $t_2<0$. (b) The honeycomb lattice with $t_1>0$ and  $t_2<0$. (c) The anisotropic triangular lattice, relevant to organic superconductors and magnets \cite{RPP}, with $t_1<0$ and  $t_2>0$. In each case two sublattices are marked; all hopping integrals are made positive by the transformation $\hat c_{i\sigma}\rightarrow\hat c_{i\sigma}$ on one sublattice only.} 
	\label{fig:lat}
\end{figure}

Thus, for example,  bipartite lattices with only nearest neighbor hopping (and the same sign of hopping between all neighbors)  are balanced. 
Some simple examples of geometrically frustrated but balanced lattices are shown in Fig. \ref{fig:lat}.

\section{Balance is sufficient for Nagaoka}\label{sect:sufficient}

%
%
%
%

In the Nagaoka limit  all states with finite energy can be written as a superposition of `single hole states' of the form
\begin{eqnarray}
|i,\mu,\tau\rangle=(-1)^{\rho(i,\mu)} \prod_{(j,\nu)\ne (i,\mu)}\hat c^\dagger_{j\nu\sigma_j}|0\rangle \label{eqn:holes}
\end{eqnarray}
where  $\tau=\{\sigma_j\}_{j\ne i}$ is a binary vector  describing the spins of all of the electrons, $|0\rangle$ is the vacuum state defined by $\hat c_{i\mu\sigma}|0\rangle=0$ for all $i$, $\mu$, $\sigma$, and $\rho(i,\mu)\in[0,L-1]$ is an arbitrary ordering of the orbitals. $\rho(i,\mu)$ need have no correlation to the physical patten of the lattice, but is required to enforce the correct antisymmetrization of the hole states -- the operators in the product are to be ordered by $\rho(i,\mu)$ with lower values to the left.

We describe two states as `directly connected' if $\langle {i,\mu,\tau}|\hat{\cal H}_t+\hat{\cal H}_X|{j,\nu,\upsilon}\rangle\ne0$. We will denote direct connection by $(i,\mu,\tau)\leftrightarrow(j,\nu,\upsilon)$.
For directly connected states
\begin{eqnarray}
\langle {i,\mu,\tau}|\hat{\cal H}_t+\hat{\cal H}_X|{j,\nu,\upsilon}\rangle=  t^*_{ij\mu\nu}. \label{Eq:expect_tX}
\end{eqnarray}

A Hamiltonian is said to satisfy the connectivity condition if there exists an integer $n$ for which  $\langle i,\mu,\tau|(\hat{\cal H}_t+\hat{\cal H}_X)^n|j,\nu,\upsilon\rangle\ne0$ for every pair of states with the same $S^z$. Notably,  one dimensional single orbital models with only nearest neighbor hopping are not connected in this sense \cite{Nagaoka,Tasaki}.

\subsection{Proof of theorem 1}

Compare the arbitrary superposition of states of single hole states
\begin{eqnarray}
|A_h\rangle=\sum_{i\mu\tau}\alpha_{i\mu\tau}|i,\mu,\tau\rangle,
\end{eqnarray}
with a fully spin polarized superposition
\begin{eqnarray}
|\Phi_h\rangle=\sum_{i\mu}\phi_{i\mu}|{i,\mu,\Uparrow}\rangle, \label{eq:gs-holes}
\end{eqnarray}
where $\tau=\Uparrow$ indicates that all the electrons are up and $\phi_{i\mu}=\sqrt{\sum_\tau|\alpha_{i\mu\tau}|^2}$. We have 

\begin{eqnarray}
\langle A_h|\hat{\cal H}_V|A_h\rangle&=&\sum_{i\mu\tau}|\alpha_{i\mu\tau}|^2 \langle i,\mu,\tau|\hat{\cal H}_V|i,\mu,\tau\rangle 
\notag \\&=&
\sum_{i\mu\tau}|\alpha_{i\mu\tau}|^2 \langle i,\mu,\Uparrow|\hat{\cal H}_V|i,\mu,\Uparrow\rangle \notag \\
&=& \sum_{i\mu}|\phi_{i\mu}|^2 \langle i,\mu,\Uparrow|\hat{\cal H}_V|i,\mu,\Uparrow\rangle 
\notag \\&=& 
\langle \Phi_h|\hat{\cal H}_V|\Phi_h\rangle.
\end{eqnarray}
That is, because $U$ is infinite and $V(\{\hat n_{i\uparrow}+\hat n_{i\downarrow}\})$  only depends on the site occupation numbers,  $\langle\hat{\cal H}_V\rangle$ is independent of the spin degrees of freedom for single hole states.

Now note that
\begin{eqnarray}
&&
\left\langle {i,\mu,\Uparrow}\left|\left(\hat{\bm{S}}_{i'\mu'}\cdot\hat{\bm{S}}_{j'\nu'}+\frac14 \hat{n}_{i'\mu'}\hat{n}_{j'\nu'}\right)\right|{j,\nu,\Uparrow}\right\rangle 
\notag\\&&\hspace*{1.5cm}
\geq \left\langle {i,\mu,\tau}\left|\left(\hat{\bm{S}}_{i'\mu'}\cdot\hat{\bm{S}}_{j'\nu'}+\frac14 \hat{n}_{i'\mu'}\hat{n}_{j'\nu'}\right)\right|{j,\nu,\upsilon}\right\rangle 
\notag \\&&\hspace*{1.5cm}
\geq 
0
\end{eqnarray} 
for all $i,i',j,j',\mu,\mu',\nu,\nu',\tau,\upsilon$. Thus, if $J_{ij\mu\nu}\leq0$ for all $i,j,\mu,\nu$ then
\begin{eqnarray}
0 \geq \left\langle {j,\nu,\upsilon}\left| \hat{\cal H}_J \right|{i,\mu,\tau}\right\rangle 
\equiv K_{i\mu\tau}^{j\nu\upsilon} \geq  K_{i\mu\Uparrow}^{j\nu\Uparrow}
\end{eqnarray}
for all $i,j,\mu,\nu,\tau,\upsilon$.
Therefore,
\begin{eqnarray}
\langle A_h|\hat{\cal H}_J|A_h\rangle
&=&
\sum_{ij\mu\nu\tau\upsilon} K_{i\mu\tau}^{j\nu\upsilon} \alpha_{j\mu\tau}^*\alpha_{i\nu\upsilon} 
\notag\\&\geq& 
\sum_{ij\mu\nu} K_{i\mu\Uparrow}^{j\nu\Uparrow} \sum_{\tau\upsilon} \alpha_{j\mu\tau}^*\alpha_{i\nu\upsilon} 
\notag\\&\geq&
 \sum_{ij\mu\nu} K_{i\mu\Uparrow}^{j\nu\Uparrow}\phi_{j\mu}^*\phi_{i\nu} 
=  \langle \Phi_h|\hat{\cal H}_J|\Phi_h\rangle,
\end{eqnarray} 
where the second inequality follows from the Cauchy-Schwartz inequality.

Because double occupancy is forbidden in the infinite-$U$ limit $\langle {i,\mu,\tau}|\hat{\cal H}_P|{j,\nu,\upsilon}\rangle=0$. 

Finally, we specialize to the case of a balanced lattice. Property \ref{2sets} of the fundamental theorem of signed graphs implies that we can construct a gauge transformation that maps the Hamiltonian onto one where all $t^*_{ij\mu\nu}\leq0$. An explicit example of such a  gauge transformation is
\begin{eqnarray}
\hat c_{i\mu\sigma}\rightarrow-\hat c_{i\mu\sigma}& \text{ for all }& (i,\mu)\in{\mathcal S}_a, \notag\\
\hat c_{i\mu\sigma}\rightarrow\hat c_{i\mu\sigma}& \text{ for all }& (i,\mu)\in{\mathcal S}_b.
\end{eqnarray}

Furthermore,
\begin{eqnarray}
&&
\langle A_h|\hat{\cal H}_t+\hat{\cal H}_X|A_h\rangle=t^*_{ij\mu\nu}\sum_{\tau\leftrightarrow\upsilon}\alpha_{j\mu\tau}^*\alpha_{i\nu\upsilon} \notag\\&&\hspace*{1.5cm}
\geq
t^*_{ij\mu\nu}\phi_{j\mu}^*\phi_{i\nu} 
=  \langle \Phi_h|\hat{\cal H}_t+\hat{\cal H}_X|\Phi_h\rangle,
\end{eqnarray}
where we have again made use of the Cauchy-Schwartz inequality and $\sum_{\tau\leftrightarrow\upsilon}$ indicates that the sum is restricted to run only over $\tau$ and $\upsilon$ such that $(i,\mu,\tau)\leftrightarrow(j,\nu,\upsilon)$, as the overlap integral vanished otherwise. Thus,  there are no states with energy lower than $|\Phi_h\rangle$.   
Theorem 1 follows immediately from the SO(3) symmetry of the model. $\square$

\subsection{Proof of theorem 2}

The Perron--Frobenius theorem \cite{Perron-Frobenius} states (among other things) that if all the elements of an irreducible real square matrix are non-negative then it has a unique largest real eigenvalue and that the corresponding eigenvector has strictly positive components. A Hermitian matrix is reducible if and only if it can be block diagonalized by a permutation matrix. Let use write the  Hamiltonian (\ref{Eq:full}) in the form
\begin{eqnarray}
\hat{\cal H}=\sum_{m=(1-N)/2}^{(N-1)/2}{\cal H}_m,
\end{eqnarray}
where $m$ labels the $z$-component of the total spin of the system. Each of the $N$ matrices $M_m=-{\cal H}_m$ is irreducible provided the Hamiltonian satisfies the connectivity condition. Furthermore, we have seen above that all of the matrix elements of ${\cal H}_m\leq0$. Therefore each of the $M_m$ satisfy the  Perron--Frobenius theorem.

The SO(3) symmetry of the Hamiltonian means that $|\Phi_h\rangle$ must be $N$-fold degenerate. As no states have lower energy than $|\Phi_h\rangle$ this means that the lowest energy states of the $S_z$-sectors are necessarily degenerate and that, up to this required   $N$-fold degeneracy, $|\Phi_h\rangle$ is unique. $\square$

\section{Frustration and the orbital part of the ground state wavefunction}\label{sec:orbital}

For the Hubbard model the explicit wavefunction can be  straightforwardly constructed.  Of course one could simply take Eq. (\ref{eq:gs-holes}) as a variational wavefunction and minimize all of the $\phi_i$. However, a more elegant approach is to introduce an ancillary model of non-interacting spinless fermions on the same lattice:
\begin{eqnarray}
\hat{\cal H}_a=-\sum_{ij\mu\nu}t_{ij}\hat c^\dagger_{i}\hat c_{j},\label{eq:non-int-e}
\end{eqnarray}
and then make a particle-hole transformation $\hat h_{i}^\dagger=\hat c_{i}$.
As this is a single particle Hamiltonian the ground state can be written in the form
\begin{eqnarray}
|\Psi_h\rangle=\sum_{i}\psi_{i}\hat h^\dagger_{i}|vac\rangle, \label{eq:non-int}
\end{eqnarray}
where the vacuum for holes, $|vac\rangle$, is the state for which $\hat h_{i}^\dagger|vac\rangle=0$ for all $i$.
Note that 
\begin{eqnarray}
|i,\Uparrow\rangle=\hat c_{i\uparrow}\prod_{j}\hat c_{j\uparrow}^\dagger|0\rangle=\hat h^\dagger_{i\uparrow}|\Uparrow\rangle
\end{eqnarray}
where $\hat h_{i\mu\sigma}^\dagger=\hat c_{i\mu\sigma}$ and $|\Uparrow\rangle=\prod_{j\mu}\hat c_{j\mu\uparrow}^\dagger|0\rangle$, where the ordering of the operators in the products is as as in Eq. (\ref{eqn:holes}). 

Recalling  Eqs. (\ref{Eq:expect_tX}) and (\ref{eq:gs-holes}) one finds that $\phi_{i}=\psi_{i}$ for all $i$, which  means that we can calculate the ground state wavefunction of the Hubbard model from the ancillary non-interaction model. Often, directly minimizing Eq. (\ref{eq:non-int}) is not the most efficient approach, for example, if the lattice is periodic a Fourier transformation leads directly to the solution.

If all $t_{ij}>0$ the ground state must have  $\phi_{i}=\psi_{i}>0$ for all $i$. That is, the wavefunction is bonding between all sites. In this sense, the ground state wavefulction is unfrustrated. Note that, in a periodic system, it always possible to construct a wavefunction that is strictly positive at high symmetry points with wavevectors, $\bm k$, satisfying $2\bm k=\bm G$, where $\bm G$ is a reciprocal lattice vector. The set of all such high symmetry points always includes the $\Gamma$-point (origin of the unit cell).

Returning to the multiorbital model,  the Perron-Frobenius theorem guarantees the existence of a gauge for which all $\phi_{i\mu}$ are strictly positive. Thus  again the ground state wavefunction is unfrustrated.

In this context it is interesting to note the recent discovery that on some topologically frustrated lattices anitferromagnetic states with magnetization near the classical limit occur in the Nagaoka limit \cite{Shastry,Arg1,Arg2}. Again here relasing the frustration  in the orbital part of the wavefunction appears to play a crucial role \cite{Arg1}.

\section{Conclusions}

We have shown that structural balance and the absence of antiferromagnetic exchange are sufficient to prove that the ground state of infinite-$U$ multiorbital model, Eq. (\ref{Eq:full}), with arbitrary pairwise interactions in ferromagnetic.

Structural balance implies the absence of topological frustration -- therefore, for itinerant electrons, balance is the natural definition of an unfrustrated lattice. While bipartite lattices (with no hopping within the sublattices) are always balanced, many non-bipartite lattices are also balanced, see Fig. \ref{fig:lat} for some examples. An interesting question, beyond the scope of this paper, is the role of structural balance to other problems involving frustration and itinerant electrons.

Balance is important because it allows for an unfrustrated orbital part of the ground state wavefunction.
This is consistent with the general insight that Nagaoka's theorem arises from the minimization of the hole's kinetic energy.

\section*{Acknowledgements}
I thank Ross McKenzie and Henry Nourse for critically reading the manuscript. 
%
This work was supported by the Australian Research Council (grants  FT130100161 and DP160100060).



\begin{thebibliography}{99}
\bibitem{foot-signs}
Note the negative sign on the right-hand-side of Eq. (\ref{Eq:tb}). Though now standard it is not included in some older papers, e.g., Refs \cite{Nagaoka,Tasaki}. This is potentially confusing when  discussing the signs of the $t_{ij}$. Note that throughout this paper all such signs are those using sign convention employed in Eq. (\ref{Eq:tb}).
\end{thebibliography}
\end{document}